\documentclass[aps,prb,twocolumn,longbibliography,footinbib,showpacs,showkeys,superscriptaddress]{revtex4-2}
\usepackage{epsfig}
\usepackage{graphicx,graphics}
\usepackage{epstopdf}
\usepackage{natbib}
\usepackage{lipsum}
\usepackage{dcolumn}
\usepackage{amsmath,amssymb,amsfonts}
\usepackage{latexsym,verbatim}
\usepackage{bm}
\usepackage{color}
\usepackage{siunitx}
\usepackage[breaklinks=true,colorlinks,citecolor=blue,linkcolor=blue,urlcolor=blue]{hyperref}
\begin{document}
	
\title{Exciton dissociation in two-dimensional transition metal dichalcogenides: excited states and substrate effects}
\author{Tao Zhu}
\affiliation{School of Electronic and Information Engineering, Tiangong University, Tianjin 300387, People's Republic of China}
\author{Chenhang Zheng}
\affiliation{Department of Applied Physics, The Hong Kong Polytechnic University, Hung Hom, Hong Kong SAR, China}
\author{Lei Xu}
\email{a0129459@u.nus.edu}
\affiliation{Department of Physics, National University of Singapore, Singapore 117551, Republic of Singapore}
\author{Ming Yang}
\email{kevin.m.yang@polyu.edu.hk}
\affiliation{Department of Applied Physics, The Hong Kong Polytechnic University, Hung Hom, Hong Kong SAR, China}
\affiliation{Research Centre on Data Sciences $\&$ Artificial Intelligence, The Hong Kong Polytechnic University, Hung Hom, Hong Kong SAR, China}
\affiliation{Research Centre for Nanoscience and Nanotechnology, The Hong Kong Polytechnic University, Hung Hom, Hong Kong SAR, China}

\begin{abstract}
Exciton dissociation plays a crucial role in the performance of optoelectronic devices based on two-dimensional (2D) transition metal dichalcogenides (TMDs). In this work, we investigate the effect of an in-plane electric field on the exciton resonance states in MX$_2$ (M = Mo, W; X = S, Se) monolayers and few-layers using the complex coordinate rotation method and the Lagrange-Laguerre polynomial expansion of the wave function. This technique enables accurate computation of both ground and excited excitonic states across a wide range of electric field strengths, overcoming limitations of previous perturbative approaches. Our calculations reveal that an electric field effectively dissociates excitons, with excited states being more easily dissociated than the ground state. The critical field for exciton dissociation is found to be smaller in WX$_2$ monolayers compared to MoX$_2$ monolayers due to the smaller exciton reduced mass. Furthermore, the presence of a dielectric substrate and an increase in the number of MX$_2$ layers enhance the exciton susceptibility to the electric field, lowering the critical field for dissociation. The dependence of exciton properties on the number of MX$_2$ layers can be well described by power functions. These findings provide valuable insights for the design and optimization of high-performance optoelectronic devices based on 2D TMDs.
\end{abstract}

\maketitle
\section{Introduction}
Two-dimensional (2D) transition metal dichalcogenides (TMDs) have garnered significant attention in recent years due to their remarkable electronic and optical properties, including sizeable direct band gaps, strong spin-orbit coupling, distinctive valley-selective circular dichroism, and robust excitonic effects \cite{1,2,3,4,xu}. The exfoliation of monolayer TMDs from their bulk counterparts induces a notable transition from an indirect to a direct band gap, transforming them into highly photoluminescent materials and promising candidates for various applications, such as field effect transistors, photodetectors, light-emitting diodes, and solar cells \cite{5,6,7,new1,new2,new3,Gogoi}. One of the most fascinating features of TMDs is the strong excitonic effects resulting from reduced screening in two-dimensional materials, which leads to the formation of tightly bound excitons with binding energies up to several hundred meV. For example, previous experimental and theoretical studies have reported large exciton binding energies ranging from 0.5 eV to 1.1 eV for monolayer MoS$_2$ \cite{4,8,9,10,11,12,new4,new5}. While these strong excitonic effects are intriguing from a fundamental physics perspective, they can hinder the performance of optoelectronic devices, as the robust binding of excitons impedes their dissociation into free charge carriers, which is essential for generating photocurrent \cite{9,11}.

The application of an external electric field has proven to be an effective method for promoting exciton dissociation in various systems, such as quantum wells \cite{14,15,16} and carbon nanotubes \cite{nano1,nano2,nano3}. Recent theoretical and experimental studies have demonstrated that applying an in-plane electric field can significantly enhance the dissociation of excitons in TMDs \cite{17,18,19,20,21,22}. For instance, Kamban and Pedersen investigated field-induced exciton dissociation in TMDs using the finite element method and exterior complex scaling, finding that the field-induced dissociation rate strongly depends on the dielectric screening environment \cite{19}. However, a comprehensive understanding of the field-induced exciton dissociation process in monolayer and few-layer TMDs is still lacking, particularly regarding the behavior of excited excitonic states and the quantitative analysis of critical dissociation field strengths with dielectric substrates.

In this work, we present a systematic theoretical study of electric field-induced exciton dissociation in monolayer and few-layer TMDs, namely MoS$_2$, MoSe$_2$, WS$_2$, and WSe$_2$. We employ the Mott-Wannier model to describe the excitons, taking into account non-local screening effects via the Keldysh potential. The density functional theory (DFT) based G$_0$W$_0$ approximation has been adopted to obtain the reduced mass and polarizability of TMDs. To solve the excitonic Schrödinger equation, we utilize the Lagrange-mesh method, which enables efficient computation of both ground and excited states of excitons. Moreover, our method effectively circumvents the singularity of the screened Coulomb potential through the use of appropriate regularized Lagrange functions. Employing this technique, we obtain a series of exciton energy levels and, notably, discover that they follow a modified Rydberg series with effective dielectric screening.

To determine the exciton resonance states under an external electric field, we employ the complex coordinate rotation method to transform the exciton Hamiltonian. This approach surpasses previous perturbation methods \cite{221,222,223} in several key aspects: it readily achieves converged exciton states even under intense electric fields, enables precise calculation of both exciton resonance energy and dissociation time, and offers superior versatility in handling multi-layer systems and substrate effects. Our comprehensive analysis of substrate influences on exciton dissociation in monolayer and few-layer MX$_2$ (M = Mo, W; X = S, Se) demonstrates that enhancing either the dielectric screening strength or the screening length significantly amplifies field-induced exciton dissociation. These results have profound implications for MX$_2$-based optoelectronic devices, suggesting that the performance of photodetectors and solar cells can be substantially improved through the strategic application of electric fields and careful selection of dielectric substrates. This work not only advances our fundamental understanding of exciton dynamics in 2D materials but also provides a robust theoretical framework for designing and optimizing future TMD-based optoelectronic devices.

\section{Method}
\subsection{Mott-Wannier model}
As a bound electron-hole pair, an exciton can be regarded as a hydrogen-like system with the hole serving as the nucleus. Since the exciton radius in 2D MX$_2$ is several times larger than the lattice constant, we describe the exciton using the Mott-Wannier model \cite{23,24}, which is appropriate for excitons with a large radius compared to the lattice spacing. In the presence of an external in-plane electric field ${\bf F}$, the Hamiltonian of the exciton in 2D MX$_2$ can be written in polar coordinates as
\begin{equation}
	H=-\dfrac{\hbar^2 }{2\mu}\nabla^2+V_{2D} (\rho)+eF\rho\cos\phi.
\end{equation}
Here, the reduced exciton mass $\mu$ is given by $\mu^{-1}=m_e^{-1}+m_h^{-1}$, where $m_e$ and $m_h$ are the effective masses of the electron and hole, respectively. $F=|{\bf F}|$ is the magnitude of the electric field strength, $\rho$ is the separation between the electron and hole in the 2D space, and $\phi$ is the angle between ${\bf F}$ and ${\bf \rho}$. For convenience, but without loss of generality, we assume the hole is localized at $\rho=0$. According to Keldysh \cite{25}, the Coulomb potential in the dielectric medium, $V_{2D} (\rho)$, due to the hole is given by
\begin{equation}
	V_{2D} (\rho)=-\dfrac{e^2}{4\varepsilon_0(\varepsilon_1+\varepsilon_2)r_0}\left[H_0 \left(\dfrac{\rho}{r_0}\right)-Y_0 \left(\dfrac{\rho}{r_0} \right)\right],
\end{equation}
where $H_0$ and $Y_0$ are the Struve function and the Bessel function of the second kind, respectively, $\varepsilon_{1}$ and $\varepsilon_{2}$ are the dielectric constants of the vacuum above and the substrate below the MX$_2$ layer, respectively. The screening length $r_0$ characterizes the spatial extent of the screening effect and is determined by the material polarizability $\chi$ through the relation $r_0=4\pi\chi/(\varepsilon_1+\varepsilon_2 )$ \cite{26}.

\begin{table}
	\caption{\label{table1}Calculated G$_0$W$_0$ band gap $E_g^{g_0w_0}$, experimental band gap $E_g^{exp}$ from literature, reduced mass $\mu$, polarizability $\chi$, and critical electric field $Ec$ for monolayer MoS$_2$, MoSe$_2$, WS$_2$ and WSe$_2$, respectively. $\gamma$ and $\nu$ are fitting parameters for effective dielectric constant.}
	\begin{ruledtabular}
		\begin{tabular}{cccccccc}
			&$E_g^{g_0w_0}$ &$E_g^{exp}$ &$\mu$ &$\chi$ &$Ec$ &$\gamma$&$\nu$\\
			&(eV) &(eV) &($m_0$)&(\AA)&(${V}/{\mu m}$)& &\\
			\hline
			MoS$_2$& 2.48 & 2.35\cite{Chiu2015} & 0.25 & 6.20 & 91 & 8.5 & 1.4 \\
			MoSe$_2$& 2.13 & 2.18\cite{10} & 0.29 & 7.13 & 85& 10.47& 1.42 \\
			WS$_2$& 2.72 & 2.59\cite{Chiu2015} & 0.18 & 6.37 & 82 & 6.21 & 1.37 \\
			WSe$_2$& 2.30 & 2.39\cite{Liu_2015} & 0.19 & 5.56 & 73& 7.11 & 1.38\\
		\end{tabular}
	\end{ruledtabular}
\end{table}
The reduced mass $\mu$ and polarizability $\chi$ can be obtained from first-principles calculations. In this work, we use the G$_0$W$_0$ approximation to calculate these quantities and the results are shown in Table I, which are in good agreement with previous studies \cite{10,Chiu2015,Liu_2015,27}. We also calculated the G$_0$W$_0$ band gaps and compared them with reported experimental values. All calculations are performed using the Vienna Ab initio Simulation Package (VASP) \cite{VASP} with the Generalized Gradient Approximation (GGA) of Perdew-Burke-Ernzerhof (PBE) exchange-correlation functional \cite{GGA}. The interactions between valence electrons and ionic cores are treated by the Projector Augmented Wave (PAW) method \cite{PAW1,PAW2}. A plane wave basis set with a cut-off energy of 500 eV is used to expand the electron wave functions. A 12$\times$12$\times$1 Gamma-centered Monkhorst-Pack \cite{MP} scheme is employed for k-point sampling in the first Brillouin zone. A vacuum layer of \SI{20}{\angstrom} is adopted in the z-direction (normal to the MoX$_2$ layer) to avoid spurious interactions between repeated slabs. The conjugate-gradient algorithm is employed for structure relaxation until the total energy converges to $10^{-6}$ eV and the Hellmann-Feynman force on each atom is less than \SI{0.01}{\electronvolt/\angstrom}.

The G$_0$W$_0$ calculations are performed with Kohn-Sham wave functions and eigenvalues to obtain the quasiparticle energies. More than 150 empty conduction bands are included with 150 eV energy cutoff for the response function throughout the calculations. These parameters have been tested to ensure convergence of the exciton binding energy within 100 meV. Moreover, it has been reported that the spin-orbital coupling (SOC) has negligible impact on exciton binding energies \cite{4}, we choose to leave out the SOC effect in our calculations to maintain consistency with the simplified Mott-Wannier model and to focus on the general trends of exciton dissociation across different TMDs.

\subsection{Complex coordinate rotation method}

When an in-plane electric field is applied to the MX$_2$ layer, the bound exciton states transform into resonance states. The complex coordinate rotation method has been successfully used to calculate atomic resonances in various applications \cite{28,29} and is well-suited for studying exciton resonance states in the presence of an external electric field. The main operation of the complex coordinate rotation method is to rotate the radial coordinate $\rho$ by a complex angle $\theta$ through the transformation $\rho\to\rho e^{i\theta}$. This transformation results in a complex rotated Hamiltonian:
\begin{equation}
	\label{eq3}
	H (\theta)=-e^{-2i\theta}\dfrac{\hbar^2}{2\mu} \nabla^2+V_{2D} (\rho e^{i\theta} )+e^{i\theta} F\rho\cos\phi.
\end{equation}

Due to the non-Hermiticity of the rotated Hamiltonian, the eigenvalues of $H(\theta)$ are complex. The radial coordinate transformation changes the divergent resonance state $e^{ik_\rho \rho}$ with $k_\rho=k-i\gamma$ ($\gamma>0$) into $e^{(-k\sin\theta+\gamma\cos\theta)\rho+i(k\cos\theta+\gamma\sin\theta)\rho}$. The state becomes convergent if $\theta >\tan^{-1} (\gamma/k)$. Generally, the eigenstates of $H(\theta)$ can be categorized into three types \cite{28}:
(i) bound states that remain unchanged under the transformation;
(ii) resonance states that are exposed from the complex plane once the rotation angle $\theta$ exceeds the argument of the complex energy of the resonance; and
(iii) continuum states that are rotated downward by an angle $2\theta$ with respect to the real axis.

We are primarily interested in the resonance states, where the complex energy is expressed as
\begin{equation}
	E=E_r-i \dfrac{\Gamma}{2}=|E|e^{i\beta},
\end{equation}
with $E_r$ denoting the energy of the resonance state and $\Gamma/2$ representing the half-width of the resonance peak, which is related to the probability of exciton dissociation. The exciton dissociation time $\tau$ is given by $\tau=\hbar/\Gamma$. $\beta$ is the argument of the complex resonance energy. When the rotation angle $\theta$ exceeds $\beta$, both the real and imaginary parts of the energy become independent of $\theta$.

\subsection{Lagrange-mesh method}
We employ the Lagrange-mesh method to obtain the eigenstates of the rotated Hamiltonian $H(\theta)$. The Lagrange-mesh method is an approximate variational method that takes advantage of convenient mesh calculations using Gaussian quadrature \cite{30}. The Lagrange-mesh method offers several advantages for our study, including efficient computation of both ground and excited exciton states, effective handling of Coulomb potential singularities, and high spectral accuracy with relatively few basis functions. These features make it particularly well-suited for our investigation of exciton properties in 2D TMDs under different dielectric environments and applied fields.

Considering that the radial variable $\rho$ ranges from 0 to $\infty$, we use Lagrange-Laguerre functions to expand the wave function \cite{31}. Accounting for the boundary conditions, the trial wave function can be expanded as:
\begin{eqnarray}
	\label{eq5}
	\nonumber\Psi(\rho,\phi)&=&\sum_{i=1}^{N}\sum_{l=0}^{L}c_{il}\varphi_{i,l}(\rho,\phi)\\
	&=&\sum_{i=1}^{N}\sum_{l=0}^{L}c_{il}f_{il}{\rho}\cos(l\phi),
\end{eqnarray}
where $c_{il}$ is the coefficient of the corresponding basis function, $i$ is the radial quantum number, $l$ is the angular momentum quantum number, and $f_{il} (\rho)$ is the Lagrange-Laguerre function defined as
\begin{equation}
	f_{il}(\rho)=(-1)^i\sqrt{\rho_i}\left(\dfrac{(N+2l)!}{N!}\right)^{-\frac{1}{2}}\dfrac{L_N^{2l}(\rho)}{\rho-\rho_i}\rho^l e^{-\frac{\rho}{2}},
\end{equation}
with $L_N^{2l} (\rho)$ being the generalized Laguerre polynomial of degree $N$ and $\rho_i$ the $i$-th zero of $L_N^{2l} (\rho)$. The Lagrange-Laguerre function $f_{il} (\rho)$ satisfies the Lagrange condition:
\begin{equation}
	f_{il} (\rho_j )=\lambda_i^{-\frac{1}{2}}\delta_{ij},
\end{equation}
and the orthogonality relation:
\begin{eqnarray}
	\label{eq8}
	\int_{0}^{\infty}f^*_{il}(\rho)f_{jn}(\rho)d\rho=A_{il}\delta_{ij,nl}.
\end{eqnarray}
Here, $\lambda_i$ is the Gauss weight associated with root $\rho_i$ and is given as
\begin{eqnarray}
	\nonumber\lambda_i&=&\dfrac{\Gamma(N+2l+1)e^{\rho_i}}{N!\rho_i^{2l+1}\left[L_N^{2l'}(\rho_i)\right]^2}\\
	&=&\dfrac{\Gamma(N+2l)e^{\rho_i}}{N!(N+2l)\rho_i^{2l-1}\left[L^{2l}_{N-1}(\rho_i)\right]^2}.
\end{eqnarray}
Moreover, $A_{il}$ in Eq.~(\ref{eq8}) is a constant which implies that $f_{il} (\rho)$ is not normalized. 

Now, the matrix elements $\langle \varphi_{il}|H|\varphi_{jn}\rangle$ of the rotated complex Hamiltonian $H$ and $\langle\varphi_{il} |\varphi_{jn}\rangle$ of the overlap matrix $S$ can be calculated in this basis set through Gauss quadrature. When there is no electric field, all Hamiltonian matrix elements will be zero for $l\neq n$. However, when $F\neq0$, the electric field couples the states $l$ and $l\pm1$, resulting in non-zero matrix elements $\langle\varphi_{il} |H|\varphi_{jl\pm1}\rangle$. Once the Hamiltonian matrix $H$ and the overlap matrix $S$ are known, we can obtain the eigenvalues and the corresponding eigenvectors, $c$, by solving the secular equation $Hc=ESc$.

\section{Results and discussion}

\subsection{Freestanding monolayer MX$_2$}
In our calculation, the wave function is expanded in a basis set of Lagrange-Laguerre polynomials with $N = 30$ and $L = 20$, which corresponds to a total of 630 basis functions. Test calculations were performed and confirmed that both the eigenvalues and the eigenvectors converge with the above choice of basis functions.

Calculations were carried out first for exciton states without the electric field. Similar to the hydrogen atom, there are many possible states, e.g., $s$, $p$, and $d$ states, in the exciton system. However, according to Fermi's Golden Rule, the probability of exciton formation by photon excitation, or the exciton oscillator strength $\Omega$, is proportional to $|\Psi(0,0)|^2$, the squared modulus of the exciton wave function at ${\bf \rho} = 0$ \cite{32}. Hence, only $s$-like exciton states are optically active due to their nonzero values of $\Psi(0,0)$ \cite{9} and are considered in this work.

\begin{figure}
	\includegraphics[width=8.6 cm]{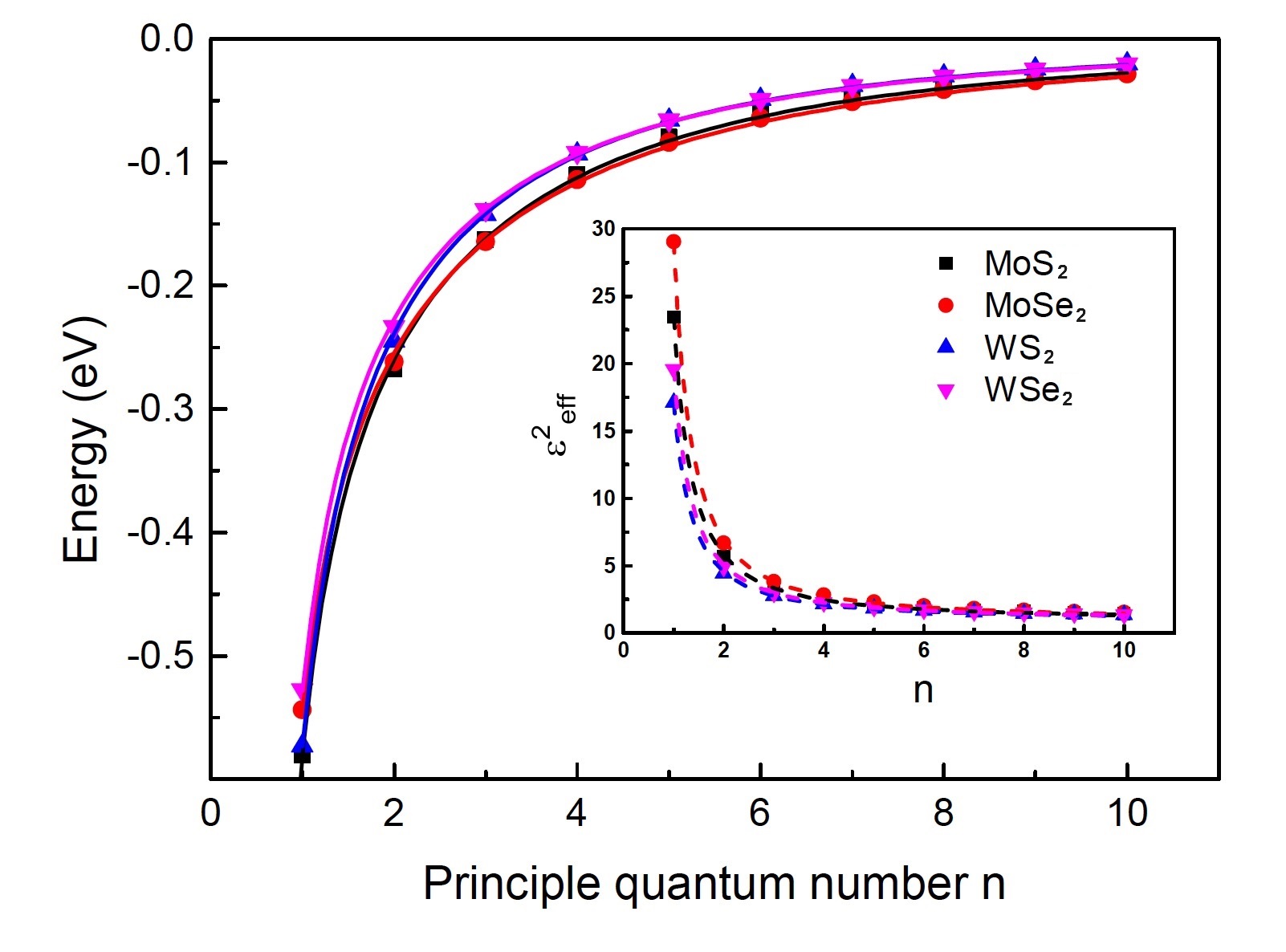}
	\caption{Energy levels for $s$-like exciton states of four monolayer MX$_2$, inset shows the fitting of effective dielectric constant with $\varepsilon_{eff}^2=1+\gamma(n-1/2)^{-\nu}$}
	\label{fig1}
\end{figure}

Fig.~\ref{fig1} shows the calculated energies of the $s$ states at different energy levels $n$ for all four monolayer MX$_2$ studied here. These energy levels are found to follow a modified Rydberg series $-\mu e^4/[2\hbar^2\varepsilon_{eff}^2(n-1/2)^2]$, but with an effective dielectric constant, $\varepsilon_{eff}$, which can be well fitted by $\varepsilon_{eff}^2=1+\gamma(n-1/2)^{-\nu}$ as shown in the inset of Fig.~\ref{fig1}. The corresponding fitting parameters are listed in Table~\ref{table1}, and it is interesting to note that the parameter $\nu$ for all MX$_2$ monolayers is very close to each other, which implies that $\nu$ could be a universal parameter for 2D materials. Physically, $\nu$ is related to the transition from the strongly screened regime at short distances to the weakly screened regime at large distances, which is characteristic of 2D materials. The near-universality of $\nu$ may suggest that this transition occurs in a similar manner across different TMDs, despite their different electronic structures. The value of $\nu$ close to 1.4 indicates that this transition occurs relatively quickly as we move to higher excited states with larger average electron-hole separations. It is also clear from the inset in Fig.~\ref{fig1} that the effective dielectric constant converges to 1 rapidly for high-lying excited states as a result of increased delocalization, revealing that they recover to the normal Rydberg series quickly as $n$ increases.

\begin{figure}
	\centering
	\includegraphics[width=8.6cm]{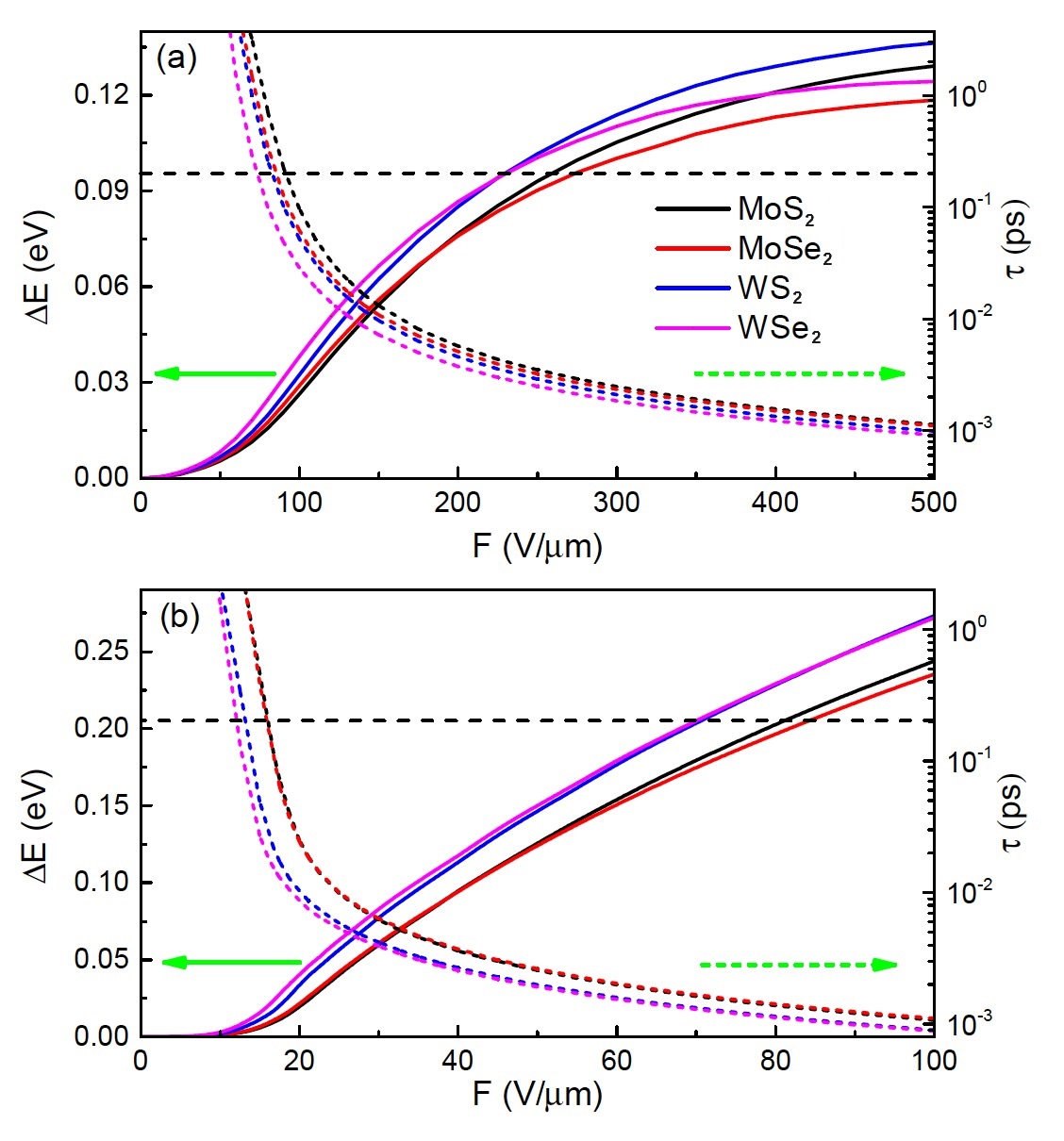}
	\caption{Energy shift $\Delta E$ (solid curve) and dissociation time $\tau$ (dashed curve) of (a) $1s$ and (b) $2s$ exciton states for monolayer MX$_2$ as a function of in-plane electric field intensity F, the dashed horizontal lines references to the smallest decay time.}
	\label{fig2}
\end{figure}

When an in-plane electric field ${\bf F}$ is applied to the MX$_2$ layer, the bound exciton states become resonance states, and a shift in energy can be expected. The calculated energy shift, defined as $\Delta E=E(F)-E(0)$, and dissociation time $\tau$ for the $1s$ and $2s$ states of the four MX$_2$ monolayers are shown in Figs.~\ref{fig2} (a) and (b), respectively, as a function of the field intensity. It is noted that the energy shift is only several meV at low electric fields, i.e., 65 V/$\mu$m for the $1s$ state and 15 V/$\mu$m for the $2s$ state. However, it increases rapidly and becomes considerable at high fields. More importantly, the exciton states become less stable and thus dissociate faster with increasing field intensity, which coincides with the exponentially decreased dissociation time shown in Fig.~\ref{fig2}. For the 1s ground state, our calculated dissociate time is in good agreement with previous reports \cite{19}. For example, for MoS$_2$ at a field strength of \SI{100}{V/\micro\meter}, our calculated dissociation time is around 10$^{13}$ s, which is exactly agree with the reported dissociation rate of 10$^{-13}$ s$^{-1}$ in Ref.~\cite{19}.

In reality, the exciton dissociation process must compete with other decay processes, such as exciton radiative and non-radiative recombination, and exciton-exciton annihilation \cite{33,34,nano1,36}. Among these decay processes, the direct radiative recombination process has been predicted to take place within a relatively smaller time scale, ranging from 0.19 ps to 0.24 ps in monolayer MX$_2$ at low temperature, and the time scale increases by one order of magnitude at room temperature \cite{34,nano1}. For exciton dissociation to dominate over the decay processes, the applied electric field should be large enough so that the dissociation time is shorter than other decay times. Using 0.2 ps as the cutoff, which is marked by the horizontal dashed line in Fig.~\ref{fig2}, the dissociation of the $1s$ exciton becomes the dominating process at the critical electric field (E$_c$) of 91 V/$\mu$m, 85 V/$\mu$m, 82 V/$\mu$m, and 73 V/$\mu$m for MoS$_2$, MoSe$_2$, WS$_2$, and WSe$_2$, respectively (Fig.~\ref{fig2}a). As shown in Fig.~\ref{fig2}b, the $2s$ state is more sensitive to the external electric field, where the same field gives rise to a larger energy shift compared to the $1s$ state. The critical field required to dissociate the $2s$ exciton state is about 6 times smaller (12-16 V/$\mu$m) compared with that for the ground state. However, the exciton oscillator strength of the $2s$ state is about 5 times smaller than that of the $1s$ state. This means that the ground $1s$ state is the main component of the photon-excited excitons, which is in agreement with the measured photoluminescence spectrum \cite{5}. Therefore, we focus on the $1s$ ground exciton state in the following discussions.

\begin{figure}
	\includegraphics[width=8.6 cm]{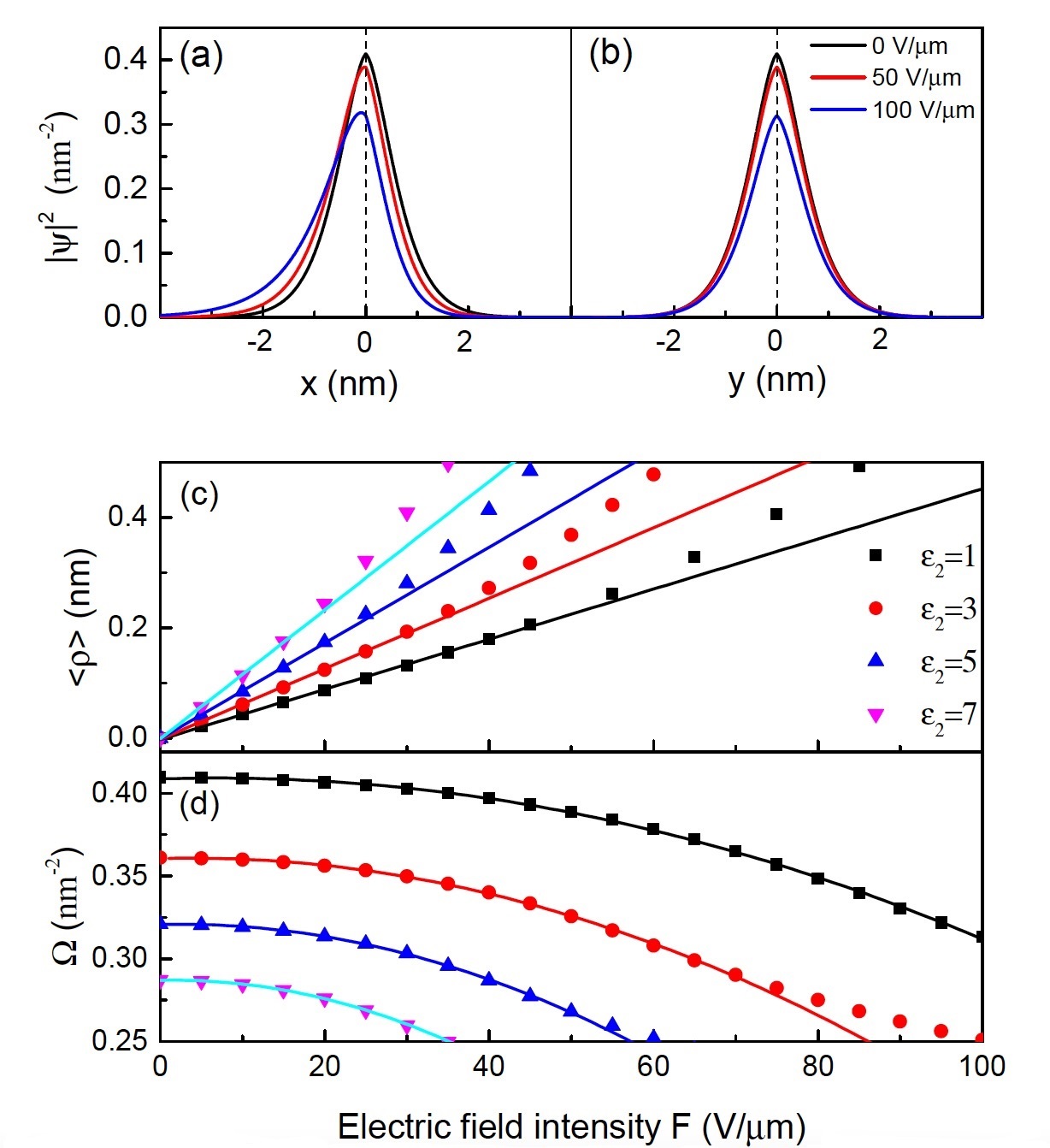}
	\caption{Distribution of monolayer MoS$_2$ $1s$ exciton wave function along (a) $x$ and (b) $y$ directions, respectively, under three different field intensities F. (c) Field induced polarization and (d) $1s$ exciton oscillator strength $\Omega$ of monolayer MoS$_2$ as a function of field intensity F in different dielectric environments.}
	
	\label{fig3}
\end{figure}

To gain more insight into the impact of the electric field on the exciton, we use the MoS$_2$ monolayer as an example to study the evolution of the exciton wave function with the external electric field. Figs.~\ref{fig3} (a) and (b) show the distribution of the exciton wave function along the $x(\phi = 0)$ and $y(\phi = \pi/2 )$ directions, respectively, in monolayer MoS$_2$ under different strengths of the field. It can be seen that the exciton wave function distributes symmetrically along the $y$-direction, but an asymmetric distribution is seen along the $x$-direction, where the exciton wave function tends to delocalize along the $-x$ direction (opposite of the electric field direction). As a consequence, the center of the electron no longer overlaps with the hole, and an electron-hole polarization is induced by the external field. The polarization can be quantified by the expectation value of the electron-hole separation $\langle \rho\rangle$, which is shown in Fig.~\ref{fig3} (c) in black squares. One can find that the polarization increases linearly with the field intensity up to 40 V/$\mu$m, and this linear dependence will result in a quadratic exciton energy shift $\alpha F^2$/2, where $\alpha$ is the slope of polarization with respect to the electric field and defined as the exciton polarizability, which can be regarded as a reflection of the sensitivity of the exciton to the external electric field. Here, for the $1s$ excitonic state of monolayer MoS$_2$, we obtained a polarizability of $4.16\times10^{-18}$ eV/(m/V)$^2$, which is in good agreement with the results obtained by the perturbation approach \cite{222} but one order of magnitude larger than the experimentally measured out-of-plane polarizability due to the localization of the exciton in the intra-layer \cite{38}. In addition, the electric field induces an apparent drop in the exciton oscillator strength, which has a quadratic dependence on the electric field over a large range of field intensity, as seen in Fig.~\ref{fig3} (d) for $\varepsilon_{2}=1$. Thus, a decrease in the intensity of the excitonic peak in the photoluminescence spectrum can be expected in the presence of an external in-plane electric field.

\begin{figure}
	\includegraphics[width=8.6 cm]{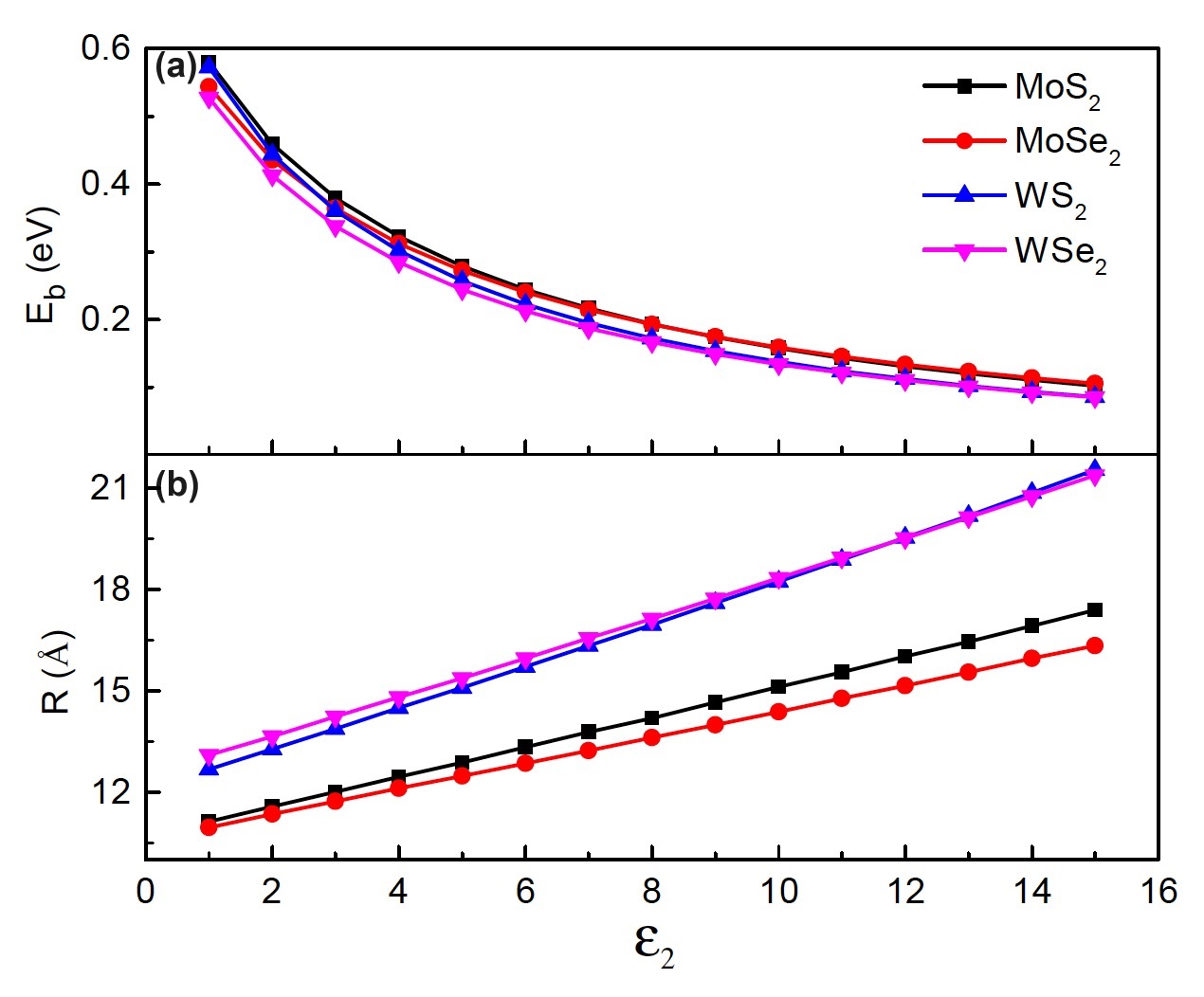}
	\caption{(a) Exciton binding energy E$_b$ and (b) exciton radius R for monolayer MX$_2$ as a function of substrate dielectric constant.}
	\label{fig4}
\end{figure}

\subsection{Substrate supported monolayer MX$_2$}

It has been reported that the substrate has a significant influence on both the electronic and optical properties of MX$_2$ \cite{10,new5,39,40}. In this section, we investigate the effect of a dielectric substrate on the field-induced exciton dissociation, and this effect is described by the dielectric constant of the substrate $\varepsilon_{2}$. The presence of a substrate will provide dielectric screening to the Coulomb interaction and result in a reduction in exciton binding energy, as shown in Fig.~\ref{fig4} (a). Meanwhile, to know the variation of the exciton wave function in the presence of a substrate, we estimate the exciton radius R of the MX$_2$ monolayers by requiring the probability of finding the electron in the hole-centered region to be 90$\%$ or more. As displayed in Fig.~\ref{fig4} (b), the radius of the exciton was found to increase linearly with the substrate dielectric constant, and the exciton in monolayer WX$_2$ is always larger than that in MoX$_2$, which is due to the smaller reduced mass of the exciton in WX$_2$ monolayer. The increase in exciton radius indicates a delocalization of the exciton wave function, which is accompanied by a decrease in exciton oscillator strength, as seen in Fig.~\ref{fig3} (d) at zero field. It is interesting to note that for the substrate-supported MoS$_2$, the oscillator strength still shows a quadratic dependence on the field intensity in a significant range of field intensity.

\begin{figure}
	\includegraphics[width=8.6 cm]{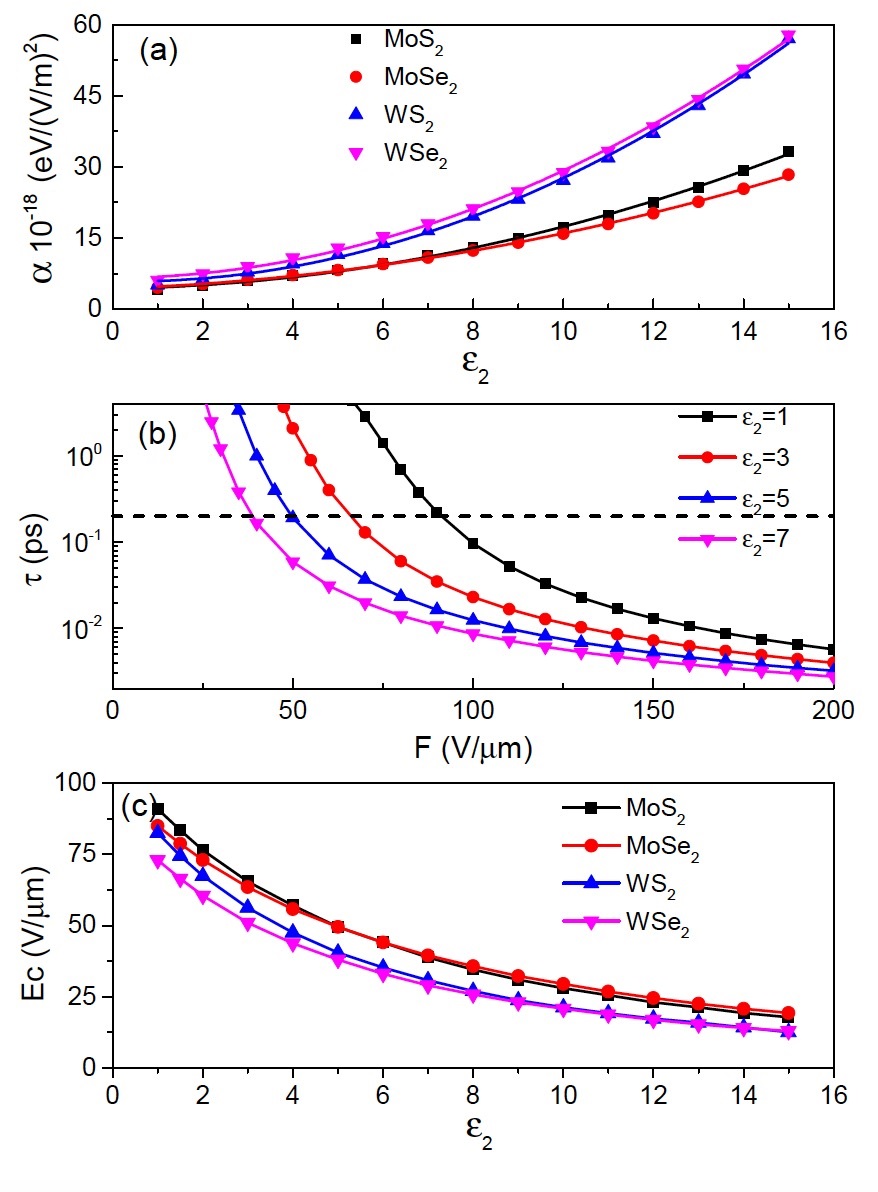}
	\caption{Substrate dielectric constant dependence of (a) exciton polarizability $\alpha$ and (c) critical electric field E$_c$ for $1s$ exciton state of monolayer MX$_2$, and the solid curve is the best fitting to a quadratic function. (b) the $1s$ exciton dissociation time $\tau$ of monolayer MoS$_2$ changes with field intensity in different dielectric environments, and the dashed line denotes the smallest decay time. }
	\label{fig5}
\end{figure}

The delocalization of the exciton wave function resulting from the increased substrate screening makes the electron-hole pair more easily separated by the electric field. As evident from Fig.~\ref{fig3} (c), the exciton polarizability increases with the substrate dielectric constant. In Fig.~\ref{fig5} (a), the calculated exciton polarizabilities for the four MX$_2$ monolayers are shown explicitly as functions of the substrate dielectric constant. It can be seen that the exciton polarizability of the WX$_2$ monolayer increases much faster than the MoX$_2$ monolayer due to the more delocalized exciton state in WX$_2$ monolayer, as discussed above. In addition, we find that the dependence of polarizability on substrate dielectric constant can be well approximated by a quadratic function.

We can now infer that the presence of a substrate makes the exciton more sensitive to the electric field, which is favorable for exciton dissociation. Indeed, Fig.~\ref{fig5} (b) shows that the exciton dissociation time decreases much faster with the applied field when monolayer MoS$_2$ is deposited on a high dielectric constant substrate, which indicates that a smaller critical electric field $Ec$ is required to induce a dissociation-dominating process. As illustrated in Fig.~\ref{fig5} (c), there is an exponential decrease in the calculated critical field with the substrate dielectric constant for the four MX$_2$ monolayers. Moreover, the reduced exciton binding energy due to the substrate screening will lead to a longer decay time \cite{41}, which implies that the substrate is more advantageous for exciton dissociation in its competition with decay processes. Therefore, a dielectric substrate can be used to assist the field-induced exciton dissociation and significantly improve the performance of electro-optical related devices, particularly photodetectors and solar cells.

\subsection{Few-layer MX$_2$}
\begin{figure}
	\centering
	\includegraphics[width=8.6 cm]{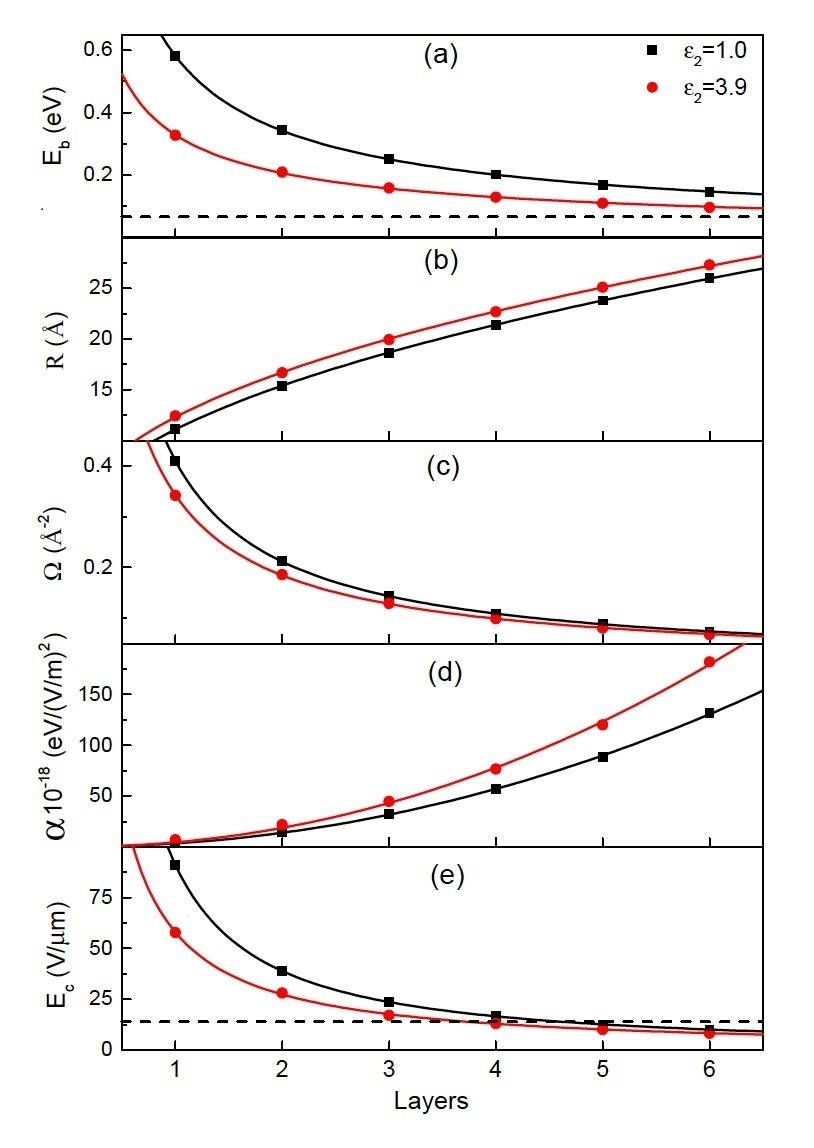}
	\caption{$1s$ exciton (a) energy E$_b$, (b) radius R, (c) oscillator strength $\Omega$ (d) polarizability $\alpha$, and (e) critical field E$_c$ of freestanding MoS$_2$ (black curve) and SiO$_2$ supported MoS$_2$ (read curve) as a function of number of MoS$_2$ layers, the dashed line in (a) and (e) represents the exciton energy and critical field of bulk MoS$_2$, respectively, whereas the solid lines are fitting curves to a power function $x^{\delta}$}
	\label{fig6}
\end{figure}

As mentioned earlier, the screening length $r_0$ of an MX$_2$ monolayer can be determined from its polarizability. In fact, we can also obtain it from $r_0=d\varepsilon/(\varepsilon_1+\varepsilon_2)$, where $d$ is the thickness of the MX$_2$ layer and $\varepsilon$ is the bulk dielectric constant of bulk MX$_2$, which suggests that the screening length has a linear relationship with the number of MX$_2$ layers. In the following, we will use MoS$_2$ as an example to investigate the effect of an external electric field on exciton dissociation in finite-layer MX$_2$. In Fig.~\ref{fig6} (a), we show the calculated exciton binding energy as a function of layer number of MoS$_2$ for both freestanding ($\varepsilon{_2}=1$) and SiO$_2$ supported ($\varepsilon{_2}=3.9$) MoS$_2$. It is clear that the binding energy initially decreases rapidly with the number of layers but slows down gradually when the number of layers is more than 3. When the number of layers becomes large, the exciton potential should recover the screened Coulomb potential in bulk form, $e^2/(4\pi\varepsilon r)$, and the corresponding exciton binding energy should approach $\mu e^4/(2\hbar^2 \varepsilon^2)$. The reduced exciton mass $\mu$, however, is similar to that in the 2D case due to its layered structure. The effective dielectric constant can be approximated as $\varepsilon=\sqrt{\varepsilon_\parallel \varepsilon_\perp}$ \cite{42}, where $\varepsilon_\parallel$ and $\varepsilon_\perp$ are the in-plane and out-of-plane dielectric components of bulk MoS$_2$, respectively, and they are estimated to be 13.8 and 5.70, respectively, by first-principles calculations. Based on these, the exciton binding energy in bulk MoS$_2$ is estimated to be 43 meV, which is indicated by the dashed line in Fig.~\ref{fig6} (a). This is in good agreement with the experimental value (50 meV) \cite{43} and the value (0.1 eV) obtained from GW-BSE calculation \cite{44}. On the other hand, along with the increased thickness, the exciton becomes more delocalized, and its oscillator strength decreases significantly due to the interlayer interaction, as shown in Figs.~\ref{fig6} (b) and (c). This drop in exciton oscillator strength could be another contributing factor to the much lower photoluminescence peak intensity observed in few-layer MoS$_2$ compared to that in monolayer MoS$_2$ \cite{45}, in addition to the direct-to-indirect band gap transition, which was believed to be entirely responsible for the lower photoluminescence intensity in finite-layer MoS$_2$.

The calculated exciton polarizabilities of MoS$_2$ are presented in Fig.~\ref{fig6} (d) for different layers of MoS$_2$. Different from the out-of-plane electric field induced exciton polarizability, which is independent of the layer thickness as measured in experiments \cite{38}, the exciton polarizability resulting from the in-plane field shows a strong dependence on the number of layers for both freestanding and SiO$_2$ supported MoS$_2$, which suggests that the exciton in few-layer MoS$_2$ is much more susceptible to the in-plane electric field. Based on the results of our calculation, we predict that the critical electric field $Ec$ required for an exciton dissociation dominating process decreases with increasing number of MoS$_2$ layers, as shown in Fig.~\ref{fig6} (e). Previous theoretical work has studied the exciton ionization in bulk MoS$_2$, and we deduce a critical field of 14 V/$\mu$m, which is highlighted in Fig.~\ref{fig6} (e), for bulk MoS$_2$ from that work \cite{46}. It seems that when the layer number reaches five, the predicted critical field is smaller than that of bulk, which is unrealistic. As the number of layers increases, inter-layer coupling becomes increasingly important and the screening becomes more complex. Our model, based on the 2D Wannier equation and Keldysh potential, does not explicitly account for these interactions. Thus, the model adopted in this work is only valid for very few layers.

\begin{table}
	\caption{\label{table2}Fitting parameters $\delta$ for exciton energy E$_b$, radius R, oscillator strength $\Omega$, polarizability $\alpha$, and critical field intensity E$_c$ shown in Fig.~\ref{fig6} with a power function $x^{\delta}$, considering frestanding MoS$_2$ and SiO$_2$ supportted MoS$_2$.}
	\begin{ruledtabular}
		\begin{tabular}{cccccc}
			&E$_b$&R&$\Omega$&$\alpha\times10^{-18}$&E$_c$\\
			&(eV)&(\AA)&(\AA$^{-2}$)&(eV/(V/m)$^2$)&(V/$\mu$m)\\
			\hline
			MoS$_2$& -0.76 & 0.48 &-0.95 &2.04 & -1.23 \\
			SiO$_2$-MoS$_2$& -0.67 &0.45 & -0.89 &2.06& -1.08\\
		\end{tabular}
	\end{ruledtabular}
\end{table}

Here, it should be emphasized that the direct band gap of monolayer MoS$_2$ becomes indirect in few-layer MoS$_2$; as a result, the proportion of indirect excitons will greatly improve with temperature. Since the indirect recombination time is several orders of magnitude larger than that of direct recombination \cite{47}, and the reduced exciton binding energy will further increase the decay time, the critical electric field obtained in this work should be regarded as an upper limit. In addition, we find that the dependence of exciton binding energy, radius, oscillator strength, polarizability, and critical electric field on the number of MoS$_2$ layers can be well fitted by a power function, and the fitted parameters are given in Table II. It is interesting to note that the exciton radius and polarizability almost show a root quadratic and quadratic dependence on the number of layers, respectively, regardless of the substrate. In contrast, the exciton oscillator strength and critical field are close to being inversely proportional to the layer number.

\section{Conclusion}
In this work, we have investigated the effect of an in-plane electric field on the exciton resonance states in MX$_2$ (M = Mo, W; X = S, Se) monolayers and few-layers using the complex coordinate rotation method combined with the Lagrange-Laguerre polynomial expansion of the wave function. The exciton properties have been well described within the Mott-Wannier model incorporating the nonlocal Keldysh potential. Our calculations demonstrate that an electric field is an effective way to dissociate excitons, with the excited states being more easily dissociated compared to the ground state. The critical field required for exciton dissociation in WX$_2$ monolayers is found to be smaller than that in MoX$_2$ monolayers due to the smaller exciton reduced mass in WX$_2$.

Moreover, we have shown that the electric field induces exciton polarization and leads to a decrease in the exciton oscillator strength, exhibiting linear and quadratic dependences on the field intensity, respectively, at moderate field intensities. Interestingly, the presence of a dielectric substrate and an increase in the number of MX$_2$ layers both enhance the exciton susceptibility to the electric field, thereby lowering the critical electric field required for an exciton dissociation dominating process. The dependence of exciton properties, such as binding energy, radius, oscillator strength, polarizability, and critical electric field, on the number of MX$_2$ layers can be well fitted by power functions. Our findings suggest that the performance of 2D MX$_2$ based electro-optical devices can be significantly improved by the application of an external in-plane electric field and the use of a dielectric substrate. The insights gained from this study provide valuable guidance for the design and optimization of MX$_2$-based optoelectronic devices, such as photodetectors and solar cells, where efficient exciton dissociation is crucial for enhanced device performance. 

Furthermore, our theoretical findings have several important implications for experimental studies of TMDs and their potential applications. First, our predicted critical fields for exciton dissociation could be tested through systematic photoluminescence (PL) quenching experiments \cite{18}. By applying in-plane electric fields of varying strengths to TMD monolayers and measuring the PL intensity, experimentalists could observe a significant drop in PL at field strengths corresponding to our calculated critical fields. This would provide a direct verification of our model. Second, our calculations of Stark shifts could be validated using electric-field-modulated reflectance spectroscopy, a technique that has been successfully applied to TMDs \cite{39}. Third, we suggest experiments where TMD monolayers are placed on substrates with varying dielectric constants. Our predictions about how the dielectric environment affects exciton dissociation could be tested by comparing the electric field required for PL quenching across these different substrates. Additionally, we propose layer-dependent studies to measure exciton dissociation in few-layer TMDs as a function of layer number \cite{1}. This could be done through layer-dependent PL quenching measurements under applied electric fields. Finally, we suggest that two-photon spectroscopy techniques may be used to probe the behavior of excited state excitons under electric fields \cite{two-photon}, potentially verifying our predictions about their easier dissociation compared to ground state excitons. These proposed experiments would not only validate our theoretical model but also provide valuable insights into the fundamental physics of excitons in 2D materials under electric fields.

\section{Acknowledgments}
T.Z. is supported by National Natural Science Foundation of China (Grant No. 12204346) and the National Key R$\&$D Program of the Ministry of Science and Technology of China (project numbers: 2022YFA1204000 and 2022YFA1402600). M.Y. acknowledges the funding support from the National Key R$\&$D Program of the Ministry of Science and Technology of China (project numbers: 2022YFA1203804), The Hong Kong Polytechnic University (project numbers.: P0034827, P0042711, P0039734, P0039679, URIS2023-050, and URIS2023-052), PolyU RCNN (Project No.: P0048122), and Research Grants Council, Hong Kong (project number: P0046939 and P0045061). The authors acknowledge the Centre for Advanced 2D Materials at the National University of Singapore for providing the high performance computing resources. 

\bibliography{exciton.bib}
\end{document}